# Label-free quantitative screening of breast tissue using Spatial Light Interference Microscopy (SLIM)

*Hassaan Majeed[1], Tan Huu Nguyen[1], Mikhail Eugene Kandel[1], Andre Kajdacsy-Balla[2], Gabriel Popescu[1,*]*

[1] Quantitative Light Imaging (QLI) Lab, Beckman Institute of Advanced Science and Technology, University of Illinois at Urbana Champaign, 405 N Matthews, Urbana, IL 61801, USA.

[2] Department of Pathology, University of Illinois at Chicago, 840 South Wood Street, Suite 130 CSN, Chicago, IL 60612, USA.

*Direct all correspondences to:     Email: gpopescu@illinois.edu
Tel: +1(217) 333-4840
Fax: +1(217) 244-1995

Breast cancer is the most common type of cancer among women worldwide. The standard histopathology of breast tissue, the primary means of disease diagnosis, involves manual microscopic examination of stained tissue by a pathologist. Because this method relies on *qualitative* information, it can result in inter-observer variation. Furthermore, for difficult cases the pathologist often needs additional markers of malignancy to help in making a diagnosis. We present a *quantitative* method for label-free tissue screening using Spatial Light Interference Microscopy (SLIM). By extracting tissue markers of malignancy based on the nanostructure revealed by the optical path-length, our method provides an objective and potentially automatable method for rapidly flagging suspicious tissue. We demonstrated our method by imaging a tissue microarray comprising 68 different subjects - 34 with malignant and 34 with benign tissues. Three-fold cross validation results showed a sensitivity of 94% and specificity of 85% for detecting cancer. The quantitative biomarkers we extract provide a repeatable and objective basis for determining malignancy. Thus, these disease signatures can be automatically classified through machine learning packages, since our images do not vary from scan to scan or instrument to instrument, i.e., they represent intrinsic physical attributes of the sample, independent of staining quality.

**Introduction:**

The latest World Health Organization (WHO) figures have reported breast cancer as the second most common form of cancer worldwide with 522,000 deaths in 2012 [1]. Within the US over 200,000 new cases of the disease are expected for women in 2017 according to the American Cancer Society [2]. Effective treatment strategies require timely and accurate diagnosis of the disease. It has been reported that, in the US, the 5-year average survival rates for patients with invasive breast cancers increase from 90% to 99% when the disease is detected at a localized (non-metastatic) stage [3].

The standard tissue evaluation method for diagnosing breast cancers involves microscopic examination of a hematoxylin and eosin (H&E) counter-stained tissue biopsy. The biopsy specimen is obtained from the patient when suspicion of disease is noted during a screening procedure such as X-ray mammography. Since cells and histological tissue sections are transparent, the H&E stain provides the necessary contrast for assessing tissue morphology using a conventional bright field microscope. This standard histopathology process has two important short-comings: reliance on qualitative markers leads to intra- and inter-observer variation while manual examination can lower the throughput of the evaluation. Quantitative microscopy could help pathologists by offering an objective assessment of the tissue physical properties. Furthermore, quantitative markers can be interpreted by machine learning classifiers for rapid analysis and automated detection [4].

In this work, we present a method for extracting quantitative markers of malignancy in breast tissue biopsies using Spatial Light Interference Microscopy (SLIM) [5]. SLIM is a quantitative phase imaging (QPI) [6] modality that generates contrast by measuring the variation of optical path-length difference (OPD) across the tissue specimen. OPD reports on the product of the refractive index and thickness of tissue at each pixel. Malignant transformation involves physical changes in epithelial cell size and density as well as the tissue organization – both of which affect OPD maps of tissue. These maps have, therefore, been used in the past for several clinical investigations [7]. This includes applications in histopathology and cytopathology including diagnosis of prostate [8] and colorectal cancers [9,10], prediction of recurrence in prostate cancer [11], analysis of Gleason grade [12], assessment of metastatic pancreatic cells [13] as well as detection of pre-malignancy in colorectal tissue [14]. Furthermore, using QPI human blood cells have also been investigated for morphological [15,16], chemical [16-18] and mechanical markers of disease [19,20].

To date, a majority of quantitative image analysis on breast tissue biopsies has relied on color images of stained tissue. Image classification in these cases has involved computing a wide range of histological features including geometric features [21,22], texture-related features [23,24] and radiometric features [23] [25,26] [see [27] for a review of methods]. However, the feature extraction process relies heavily on tissue staining which can vary from sample to sample and instrument to instrument, affecting the robustness of the classifier [28]. The label-free approach we propose makes classification through machine learning easier since the instrument does not require calibration for inconsistency in pixel values due to variations in staining, tissue changes caused by harsh solvents etc. Other label-free quantitative methods for tissue image classification have been proposed in the literature, including Fourier transform infrared spectroscopy (FTIR) [29-31], Raman spectroscopy [32-34], optical coherence tomography (OCT) [35,36] and second-harmonic generation (SHG) imaging [37,38]. However, these techniques differ from our QPI-based method in terms of speed, resolution, and compatibility with the current diagnostic pipeline.

We demonstrated in our previous work [39] that SLIM captures sufficient tissue morphology to separate benign from malignant tissue via visual investigation by trained pathologists. In this work, we demonstrate the quantitative analysis capabilities of our tissue screening system by imaging a tissue microarray (TMA) comprising 68 different cases (34 benign and 34 malignant). For each epithelial region within a tissue core, we extracted scattering, geometric, and texture-related markers of tissue malignancy from the SLIM maps (see Materials and Methods). A linear-discriminant analysis (LDA) classifier was trained to separate benign cases from malignant cases and three-fold cross validation was performed to measure the classification accuracy of the learned model [40,41]. Using validation by the Receiver Operating Characteristic (ROC) curve analysis, our results revealed a sensitivity of 94% and specificity of 85%.

**Materials and Methods**

a. **SLIM Optical Setup**

Figure 1 illustrates the SLIM optical setup which has been discussed in detail in previous publications [5,42]. The setup comprises of a module (CellVista SLIM Pro, Phi Optics, Inc.) coupled to the output port of a commercial phase contrast microscope (Carl Zeiss, Axio Observer Z1). This compatibility with existing microscopes promises to reduce barriers to clinical adoption since optical microscopes are commonly available in pathology labs. In the SLIM module, the conjugate image plane outside the microscope is relayed onto a CCD camera (Andor, Zyla) using a 4f system comprising lenses $L_1$ and $L_2$. At the Fourier plane of $L_1$, a spatial light modulator (Boulder Nonlinear Systems) is used to modulate the phase difference between the scattered and unscattered components of light in increments of $\pi/2$. Four different modulations are applied [Fig. 1 (b)] and the resulting phase image is reconstructed using a previously published algorithm [5]. Using a software platform developed in-house, the SLIM module has been upgraded with full-slide scanning capabilities [9,39]. The acquisition speed is in the range of the existing commercial tissue scanners, which, in turn, only perform bright field imaging [39]. Throughout our experiments, a 40x/0.75 NA phase contrast objective was used for imaging.

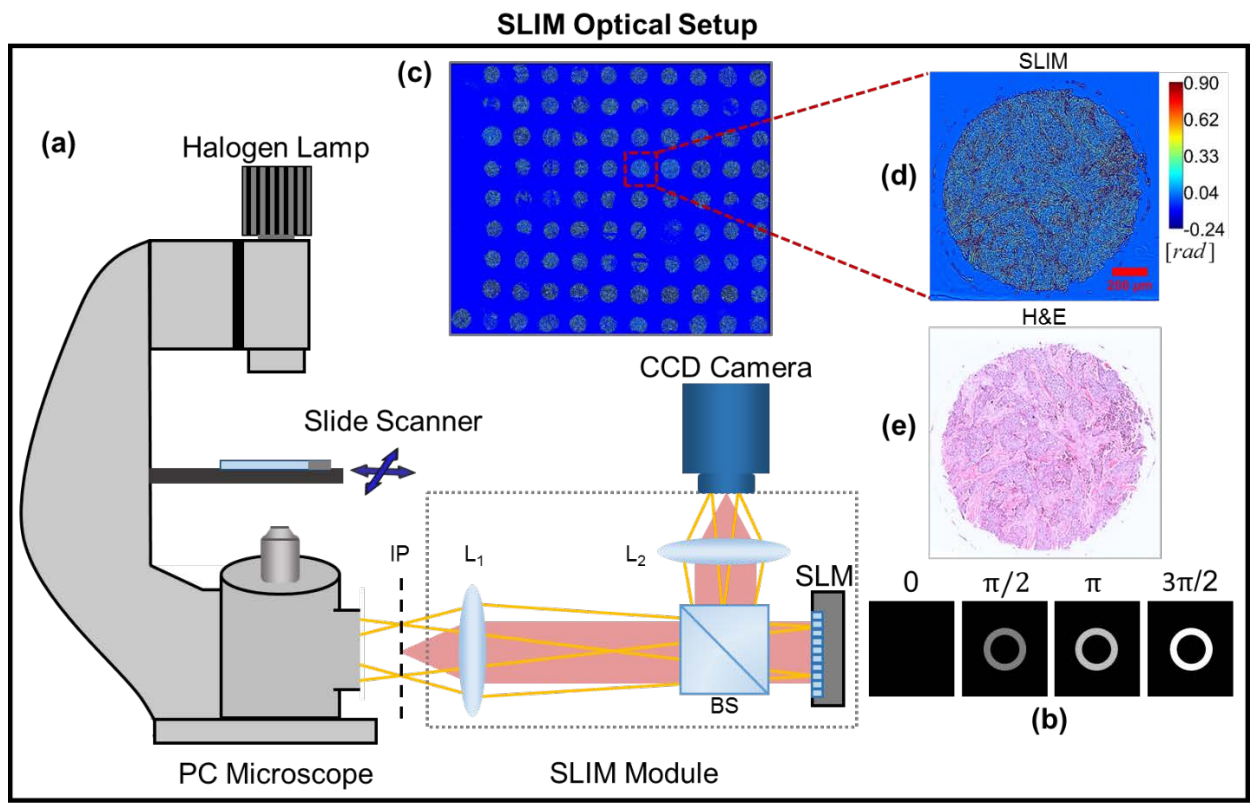

**Figure 1. (a)** The SLIM module added on to a commercial phase contrast microscope. **(b)** Four frames are acquired to compute one phase image by modulating the phase difference between scattered and incident light using a spatial light modulator (SLM). **(c)** An image of the whole slide scanned using SLIM. **(d)** Example of a TMA core SLIM image. **(e)** Bright field image of the same core after H&E staining. BS, beam splitter; $L_1$-$L_2$, lenses; IP, image plane.

**b. Tissue microarray**

The TMA used for our study was purchased from US Biomax Inc. (Serial # BR-1002) with diagnosis for each case provided by the manufacturer. The TMA was obtained with all human subject information de-identified. Neither the authors of this work nor their institutions were involved in the collection of tissue. The TMA comprised of cores 1 mm diameter and a section thickness of 5 µm. Standard formalin fixation and paraffin embedding (FFPE) histological preparation was used for each tissue block before extraction of cores. A xylene based mounting medium was used during cover-slipping.

The TMA consisted of 36 cases of infiltrating ductal carcinoma (IDC) and 36 cases of tumor adjacent normal tissue (one core per case – a total of 72 cores). Three of the tumor adjacent normal cores were obtained from the IDC cohort. In addition, 10 normal cases were included (single core each) from autopsy procedures. For final analysis we selected 34 cores diagnosed as malignant and 34 cores diagnosed as normal (either tumor adjacent normal or normal). The selection criteria were based on whether the core was intact and whether any epithelial tissue was present in the core (cores containing only stromal tissue were excluded). A SLIM image of the whole TMA slide is illustrated in Fig 1 (c) while Figs. 1 (d) and (e) show, respectively, phase map and H&E stained tissue bright field image (henceforth referred to as 'H&E image') of one core. For obtaining a mosaic of the TMA, we used a C++ based stitching code, developed in-house [9]. After staining the same tissue slide using standard protocols [43], H&E images of the TMA were acquired using a bright-field microscope (Carl Zeiss, Axio Observer Z1) outfitted with a color camera (Carl Zeiss, Axiocam MRC). The H&E images were used to assist with annotation of epithelial regions in tissue, discussed below.

c. **Annotation of epithelial regions in tissue images**

Glands or continuous epithelial regions within each core were manually annotated using the region of interest (ROI) tool of ImageJ to allow feature extraction for each gland. A consistent criterion for annotation was used where groups of epithelial cells bounded by stroma on all sides where considered a single gland. Other tissue components within epithelium (such as lumen etc.) were considered part of the gland if bounded on all sides by epithelial cells. Glands from cores in the IDC cohort were labelled as malignant while those from cores in the tumor adjacent normal cohort were labelled as benign.

d. **Extraction of geometric and scattering features**

Malignant transformation in breast tissue affects the size, shape and density of epithelial cells as well as the shape and organization of epithelial tissue. As a result, both the geometry and scattering properties of the gland are affected. We used gland perimeter curvature $C$, as well as the mean scattering length $l_s$ as part of the feature set used for separating benign and malignant tissue. The parameter extraction process is illustrated in Fig. 2 and a detailed description for each is provided below.

The extrinsic curvature $C$ of a two-dimensional plane curve $P(x, y)$, that is parametrized by Cartesian coordinates $x(t)$ and $y(t)$ with parameter $t$, is given by the expression [44]

$$C(t) = \frac{|x'y'' - y'x''|}{\left(x'^2 + y'^2\right)^{\frac{3}{2}}}, \qquad (1)$$

where the $x'$, $y'$ and $x''$, $y''$ refer to the first and second derivatives in $t$, respectively. In the above parametrization, $t$ refers to each pixel comprising the curve $P(x, y)$, having coordinates $x(t)$ and $y(t)$. This curvature can be interpreted as the magnitude of the rate of change of a vector tangent to $P(x, y)$. We computed $C$ for the perimeter $P(x, y)$ of each annotated gland by using an open source MATLAB code [45]. The code approximates $P(x, y)$ as a polygon before computing $C$ for each point defining the gland perimeter, as described in Eq. (1). To speed up computation, the image of each core was first down-sampled from the raw image size of 8000 x 8000 to 2048 x 2048 pixels. The perimeter $P(x, y)$ was then further down-sampled by a factor 20 before computing $C$ in order to remove any pixel level errors due to manual annotation. The median gland curvature $\langle C \rangle$ was then used as a feature for separating benign and malignant cases. Figs. 2 (c) and (d) illustrate $\langle C \rangle$ for representative benign and malignant glands.

The mean scattering length $l_s$ is a bulk scattering parameter that defines the length scale over which a single scattering event occurs on average. Assuming that the tissue slice captures the refractive index spatial fluctuation statistics, i.e., assuming statistical homogeneity, $l_s$ can be computed through the *scattering-phase theorem* using the expression [46]

$$l_s = \frac{L}{\text{var}[\phi(x, y)]}, \qquad (2)$$

where $\phi(x, y)$ is the SLIM phase image, $L$ is the tissue section thickness and the operator $\text{var}[.]$ computes the spatial variance over a region. The $l_s$ parameter has been used in the past for discriminating between benign and malignant prostate tissue [8]. We first computed the image $l_s(x, y)$ using a variance filter kernel size of 149 x 149 pixels, which equals the approximate diameter of 3 epithelial cells. The feature $\langle l_s \rangle$ was then computed by calculating the median of $l_s(x, y)$ over the gland area. This computation is illustrated in Figs. 2 (e) and (f).

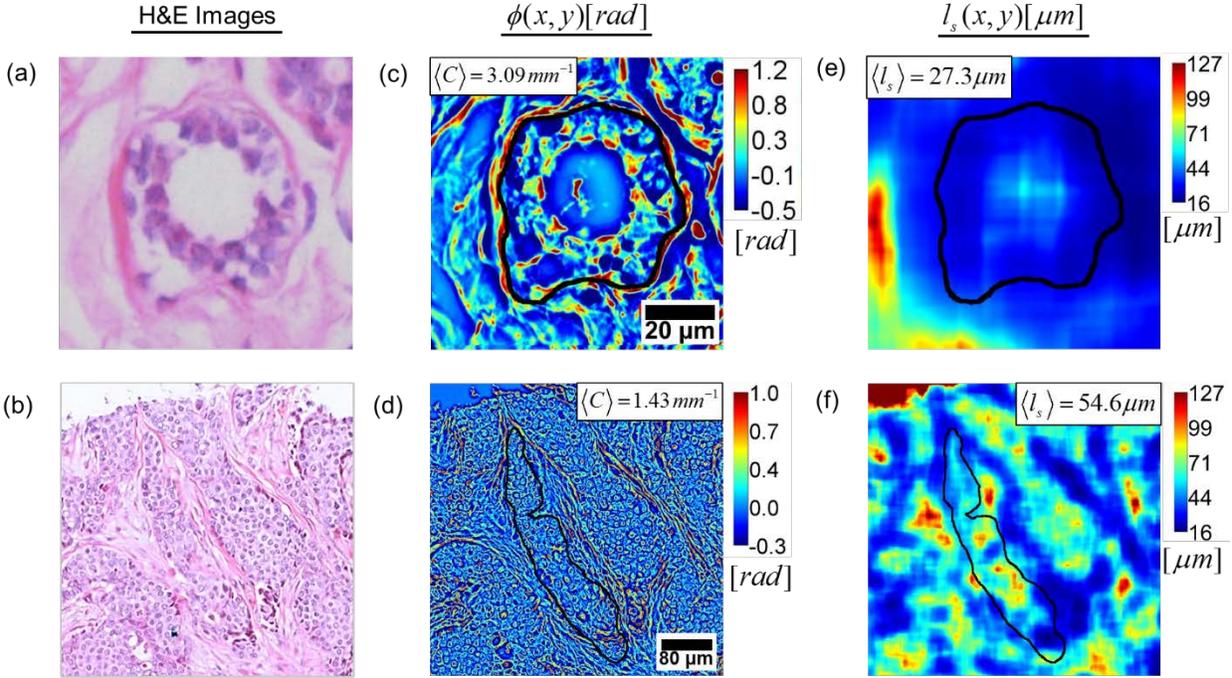

**Figure 2.** Computing the geometric feature $\langle C \rangle$ and scattering feature $\langle l_s \rangle$ over each annotated gland. **(a)** and **(b)** H&E images of benign and malignant glands, respectively. **(c)** and **(d)** SLIM images of the same benign and malignant glands, respectively, illustrating gland curvature $C$. The median over gland $\langle C \rangle$ is used as the geometric feature for classification. **(e)** and **(f)** $l_s(x, y)$ for benign and malignant glands, respectively. The median over gland $\langle l_s \rangle$ is used as the scattering feature for classification.

e. **Extraction of texture-related features**

Benign and malignant epithelial tissues differ not only in cell morphology but also in the organization of their components, leading to different textures. Texture-related features have been used in the past for solving different classification problems in histopathology of cancers [12,27]. Our feature extraction follows the work done by Varma *et al.* [47] for classifying different materials based on their texture. The approach is illustrated in Fig. 3. A TMA core phase image [Fig. 3 (a)] is first filtered through a convolution with the Leung-Malik (LM) filter bank [Fig. 3 (b)]. This filter bank consists of gradient filters (both odd and even) at different orientations and spatial scales [48]. In total, 58 different filters were used, generating a 58-dimensional response vector for each pixel in the core phase image [Fig. 3 (c)]. K-means clustering was then performed on the response vectors (number of clusters, K = 50) generated from all cores within the training set and the computed cluster centroids were referred to as 'textons' [47,48]. Since each pixel in the core belongs to a texton, for each pixel the histogram of textons was generated for its vicinity (window size 60 x 60 pixels) and was used to characterize the local texture in that neighborhood. This way, a 50 dimensional feature vector $T$ was generated to characterize texture in a pixel's neighborhood. An open source MATLAB code was used for generating the LM filter bank for this work [49].

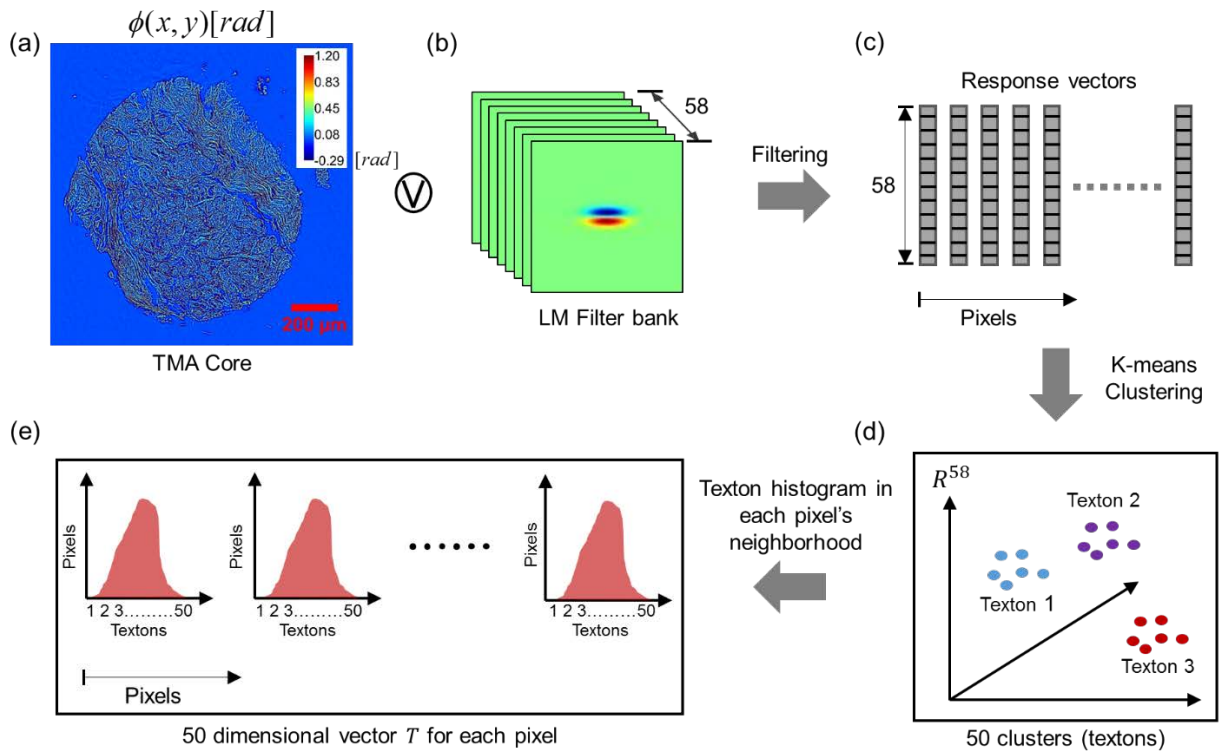

**Figure 3.** Algorithm for computing the texture in a pixel's neighborhood. **(a)-(c)** Generating the response of each pixel to a Leung-Malik filter bank. **(d)** K-means clustering of response vectors, generated from all cores in the training set, in order to find 50 cluster centroids or textons. **(e)** Histogram of textons, within a pixel's neighborhood, comprise the texture-related feature vector $T$ for each pixel.

**f. Classifier training and validation**

Since our work involves classifying each gland within a tissue core as benign or malignant, a feature vector for each gland was next generated by concatenating geometric, scattering and texture-related features. This procedure is illustrated in Fig. 4. After pixel-wise computation of gland curvature $C$, scattering length $l_s$ and texture vector $T$, the median of each feature was computed over each gland in a core and a combined 52 dimension feature vector was generated for training. For each gland, this feature vector was then used as a predictor for training a linear-discriminant analysis (LDA) classifier [Fig. 4(a)]. Class labels, either benign or malignant, were used as the ground-truth for each gland during the training process.

The feature extraction for validation purposes, illustrated in Fig. 4 (b), followed a nearly identical procedure to that used during training. The only difference was that, instead of finding new textons (cluster centroids) for validation data, the texture feature vector $T$ was computed by using the same textons as determined during training. As in training, a 52 dimensional feature vector was input to the LDA classifier which then used the model learned during training to generate a likelihood score for a gland being benign or malignant. Finally, the mean of the likelihood scores of all glands within a core was computed and used as the likelihood score of a core being benign or malignant. These scores were then used to generate a receiver operative characteristic (ROC) to select an operating point for separating benign and malignant cases (see Results and Discussion).

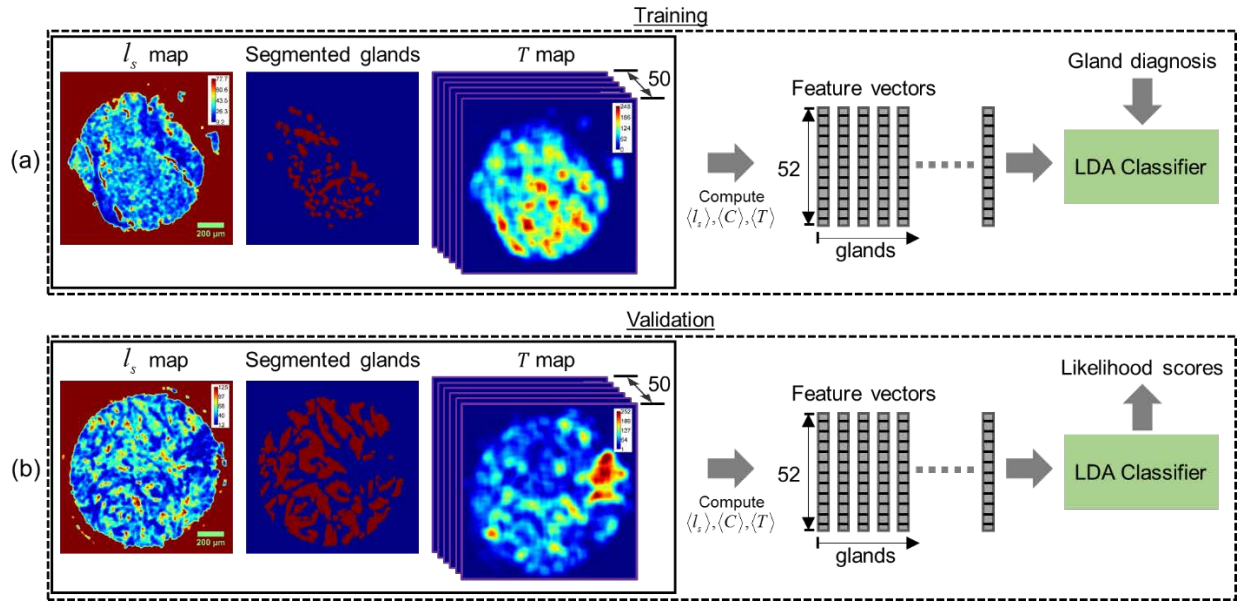

**Figure 4. (a)** Training and **(b)** Validation procedure for classifying glands as benign or malignant.

**Results and Discussion**

The classification results of our analysis are summarized in Fig. 5. In order to evaluate the accuracy of our method, we performed three-fold cross-validation [50] as illustrated in Fig. 5 (a). The total number of cases were divided into three (nearly) equal groups. In each trial, two groups were used for training while the remaining one was used for validation. Thus, three validation trials were performed, each time selecting a different validation/training set combination.

Figure 5 (b) illustrates the separation between benign and malignant gland feature vectors in one training set. In order to illustrate the data separation in 3 dimensions, we use principal component analysis (PCA) and represent the 50-dimensional feature vector $\langle T \rangle$ through its first principal component PC1 $\langle T \rangle$. The training space shows that scattering feature $\langle l_s \rangle$ has on average higher values for malignant glands than for benign glands. This finding is compatible with typical gland morphology in breast tissue since benign glands are well differentiated, consisting of a number of different structures including epithelial cells, lumen and myoepithelial cells [51]. This heterogeneity of structure results in short mean scattering lengths as explained by a large variance in Eq. (2). Malignant glands on the other hand consist of a monoclonal proliferation of cells, sometimes even showing sheets of poorly differentiated epithelial cells, resulting in smaller variance and larger $l_s$ values [51]. These phenomena can also be observed in the examples given in Figs. 2 (e) and (f). In previous investigations on prostate cancer, it was shown that $l_s$ has a lower value in malignant tissue than in benign tissue [8]. That analysis, however, was carried out on larger areas of tissue where cellular organization can be different from the epithelial only regions we are studying in this work [8].

The median gland curvature $\langle C \rangle$, on the other hand, generally has higher values for benign glands than for malignant glands. This is a result of the fact that the edge of a benign gland is constrained to follow a round or elliptical shape due to tubule formation [Figs. 2 (a) and (c)] [51]. When malignant transformation occurs, this constraint is broken and the gland edge is more irregular. At the spatial scale of investigation we have used here (approx. 13 $\mu m$), the perimeter of the malignant gland is less rapidly varying, on average, than that of a benign gland. This geometric feature is similar to the previous measurement of the gland perimeter fractal dimension that has been used for histopathology [21,27].

Fig. 5 (c) shows the separation between benign and malignant glands in the validation feature space, where, qualitatively, the same separation trend is seen as in training. We show the results of only one of the three validation trials that were carried out. As described in Materials and Methods, the gland likelihood scores, generated by the classifier during validation, were averaged over each gland in order to obtain core-wise or case-wise scores. The core-wise likelihood scores from the 3 trials were then pooled together to generate the ROC curve illustrated in Figure 5 (d) [52]. Our results indicate an area under the curve (AUC) of 0.91. The optimum operating point for classification was determined by assigning equal weight to the cost of misclassifying positives and the cost of misclassifying negatives [53]. This resulted in a sensitivity of 0.94 and specificity of 0.85 for the three-fold cross validation. The higher sensitivity of 0.94 is appropriate for our screening application, as a diagnostic tool based on this method will generate a small number of false negatives. Such a tool can point out tissue areas that have even a small chance of malignancy so a pathologist can inspect them more closely.

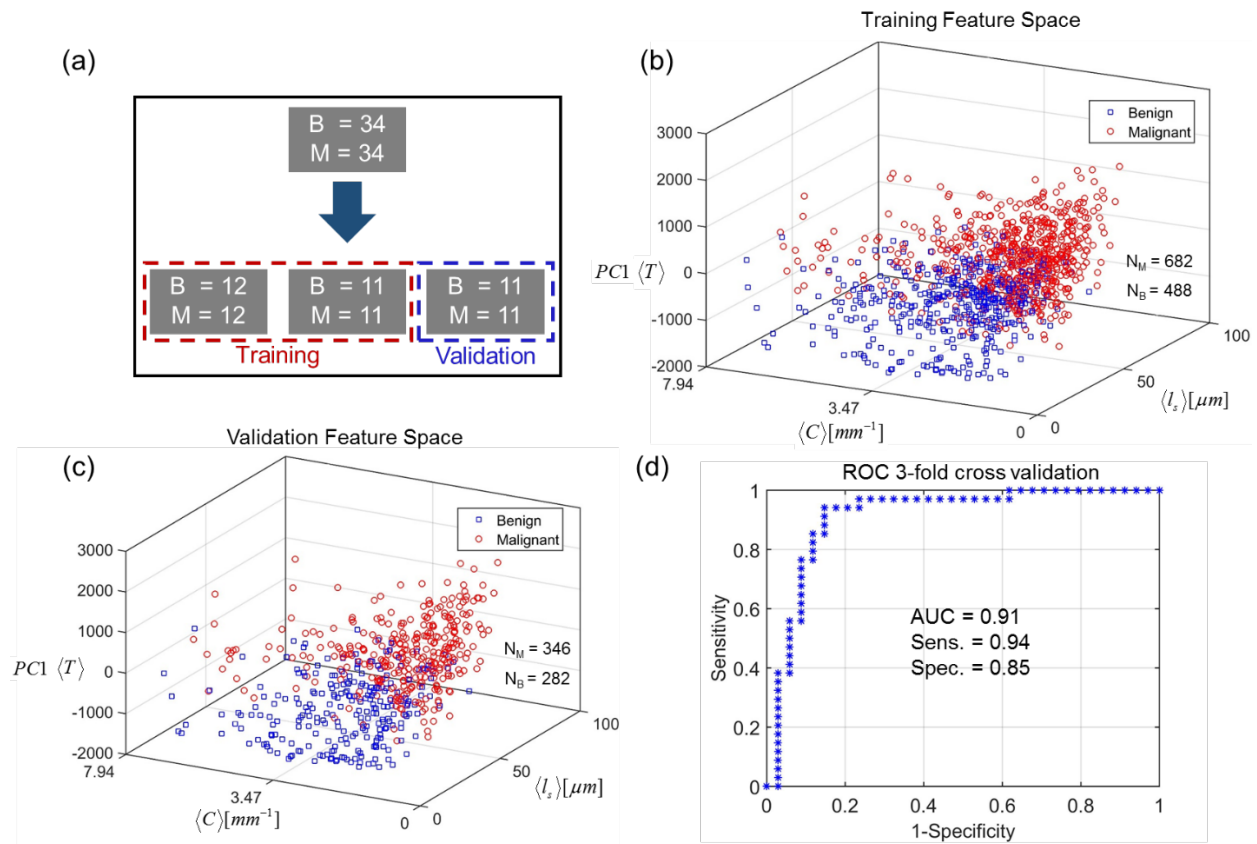

**Figure 5. (a)** Three-fold cross-validation procedure for evaluating classification accuracy. **(b)** Separation of benign and malignant gland feature vectors during *training* in 1 of 3 validation trials. **(c)** Separation of benign and malignant gland feature vectors during *validation* in 1 of 3 validation trials. **(d)** ROC curve for the 3 validation trials resulting in a sensitivity of 0.94 and specificity of 0.85 at the optimum operating point.

**Summary and Conclusions**

In summary, we presented a new method for screening tissue biopsies obtained from patients under investigation for breast cancer. Since our method relies on measurement of OPD maps, an intrinsic property of tissue, the basis for classification is objective and not subject to inter-observer variation. While in the past much of the quantitative histopathology has relied on analysis of stained tissue, our method performs image processing and machine learning on unlabeled images, making it insensitive to variability due to staining. Thus, the process of automating the entire method is feasible and subject to our future efforts.

While other label-free diagnosis methods have been proposed for these types of investigations, they affect the standard diagnostic pipeline in terms of either speed, resolution or compatibility with established workflow. SLIM, on the other hand, requires minimal changes to a conventional microscopic optical train due to its modular design. Equipped with a slide-scanning feature for rapid acquisition, a SLIM tissue scanner can potentially carry out high-throughput automated histopathology, not only reducing the case-load for pathologists but also providing complementary information through new markers. This carries the potential for incorporation into daily practice of diagnostic surgical pathology, either as a screening method to point out areas of the slide that need additional attention, or for difficult cases where pathologists need supporting tests to make a final diagnostic decision.

**Acknowledgements:**
We would like to thank Kingsley Boateng for his help with the H&E staining procedure for the tissue biopsies. This work was supported by the National Science Foundation (CBET-1040461 MRI, CBET-0939511 STC, DBI 1450962 EAGER and IIP-1353368). H. M's PhD research is being sponsored by the Beckman Graduate Fellowship Program administered through the support of the Arnold and Mabel Beckman Foundation.

**Competing financial interests:**

G. P has financial interest in Phi Optics, Inc., a company that develops quantitative phase imaging technologies.

**Author contributions statement:**

H. M imaged the tissue samples and performed the image processing and analysis. He also wrote the main manuscript text and prepared the figures. T. N helped with the image processing and analysis while M. K developed the tissue scanner used for imaging the samples. A. B provided inputs from a clinical perspective and G. P supervised the project.

**Data availability statement**

The datasets generated or analysed during this work are available from the corresponding author on reasonable request.

**Figure Legends:**

**Figure 1.** (**a**) The SLIM module added on to a commercial phase contrast microscope. (**b**) Four frames are acquired to compute one phase image by modulating the phase difference between scattered and incident light using a spatial light modulator (SLM). (**c**) An image of the whole slide scanned using SLIM. (**d**) Example of a TMA core SLIM image. (**e**) Bright field image of the same core after H&E staining. BS, beam splitter; $L_1$-$L_2$, lenses; IP, image plane.

**Figure 2.** Computing the geometric feature $\langle C \rangle$ and scattering feature $\langle l_s \rangle$ over each annotated gland. (**a**) and (**b**) H&E images of benign and malignant glands, respectively. (**c**) and (**d**) SLIM images of the same benign and malignant glands, respectively, illustrating gland curvature $C$. The median over gland $\langle C \rangle$ is used as the geometric feature for classification. (**e**) and (**f**) $l_s(x,y)$ for benign and malignant glands, respectively. The median over gland $\langle l_s \rangle$ is used as the scattering feature for classification.

**Figure 3.** Algorithm for computing the texture in a pixel's neighborhood. (**a**)-(**c**) Generating the response of each pixel to a Leung-Malik filter bank. (**d**) K-means clustering of response vectors, generated from all cores in the training set, in order to find 50 cluster centroids or textons. (**e**) Histogram of textons, within a pixel's neighborhood, comprise the texture-related feature vector $T$ for each pixel.

**Figure 4.** (**a**) Training and (**b**) Validation procedure for classifying glands as benign or malignant.

**Figure 5. (a)** Three-fold cross-validation procedure for evaluating classification accuracy. **(b)** Separation of benign and malignant gland feature vectors during *training* in 1 of 3 validation trials. **(c)** Separation of benign and malignant gland feature vectors during *validation* in 1 of 3 validation trials. **(d)** ROC curve for the 3 validation trials resulting in a sensitivity of 0.94 and specificity of 0.85 at the optimum operating point.